# METHOD FOR MEASUREMENT OF THE GRID TRANSPARENCY IN IONIZATION CHAMBERS

A. A. VOROBYEV, V. A. KOROLEV

Leningrad Institute of Physics and Technology

**Abstract**

A method is proposed for measurements of the transparency coefficient of the grids used in ionization chambers. The method is based on application of the Green's reciprocity theorem and gives an accuracy of (3 – 5) %. The experimental data are compared with the results of theoretical calculations [1] for the case of flat grids.

**Introduction**

As known, in order to reduce the induction effect of positive ions on the anode of impulse ionization chambers, a screening grid is introduced between the cathode and the anode. In this case, the magnitude of the pulses arising on the collecting electrons anode is determined by the expression :

$$V = -N \cdot e/C + V^+, \qquad (1)$$

where $N$ is the number of the ion pairs formed during the passage of the ionizing particle in the working volume of the chamber between the cathode and the grid; $e$ is the absolute value of the electron charge; $C$ is the capacitance of the anode relative to the ground; $V^+$ is the potential arising on the anode under the action of the chain of positive ions located between the cathode and the grid. Since the value of $V^+$ depends on orientation of the track, this leads to deterioration in the resolution of the ionization spectrometer.

A theoretical consideration of the grid transparency was presented in ref. [1]. These calculations, based on the method of conformal transformations, are rather complicated and, in addition, they are performed with some simplifying assumptions. So far, no experimental verification of the calculations has actually been carried out. A method for the experimental determination of the grid transparency coefficient in the ionization chamber is described below, and the experimental data are compared with the calculation results from [1].

**The experimental method for determination of the $V^+$ potential**

When determining the potential $V^+$ arising on the anode under the action of the positive ions, we can use the Green's reciprocity theorem [2]. First, we determine the potential induced on the anode by the i-th positive ion located at the distance x from the cathode (Fig. 1). Using the Green's theorem, one can show that

$$V^+_i = e \cdot \phi(x) / C \cdot U \qquad (2)$$

where $\phi(x)$ is the potential at the distance $x$ from the cathode which arises under the following conditions: the cathode and the grid are grounded, the anode is at potential $U$, and any charge in between the chamber electrodes is removed from the system.

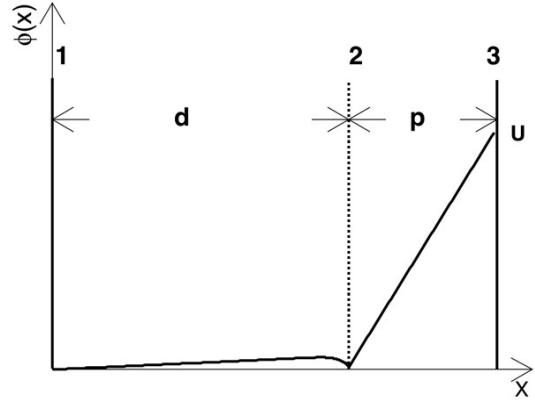

**Fig. 1.** *The potential distribution in the chamber when determining the grid transparency coefficient.* 1 - cathode, 2 - grid, 3 – anode.

The potential distribution for this case is shown in Fig.1. We are interested in the region between the cathode and the grid. At a sufficient distance from the grid, the field $E_d$ is constant. This is true when the distance from the grid exceeds ~ 5 $S$, where $S$ is the wire spacing in the grid. Then we have

$$\phi(x) = E_d\, x. \qquad (3)$$

To determine $V^+$, one needs to sum up expression (2) over all positive ions:

$$V^+ = N \cdot e \cdot E_d \cdot R \cdot \cos\Theta / C \cdot U \qquad (4)$$

where $R$ is the average distance of the ions from the beginning of the track, $\Theta$ is the angle between the direction of the track and the normal to the surface of the electrodes. We consider here the ionization created by alpha-particles emitted from a source deposited on the cathode. Thus, the task to determine the effect of positive ions on the anode came down to determination of the field strength $E_d$ in the above described conditions. As we shall show, the value of $E_d$ can be easily obtained experimentally. For this, the chamber is set into the following mode. The chamber is filled with pure argon. A sufficiently high voltage of U = + 3 kV is applied to the anode, while the grid and the cathode remain at the zero potential. The pressure in the chamber is set so that the alpha-particles stop at such a distance from the grid where the field is still constant. The field $E_d$ penetrating through the grid moves a part of the electrons formed during ionization towards the grid, and the pulses appear on the anode. Then, a compensating positive voltage $\Delta U$, at which the pulses disappear, is applied to the cathode. The disappearance of the pulses means equality of the compensating field $E'_d$ and $E_d$. The value of $E'_d$ can be determined via the value of the compensating voltage $\Delta U$ at the cathode :

$$E'_d = E_d = \Delta U/d, \qquad (5)$$

where $d$ is the distance between the cathode and the grid. Strictly speaking, it would be necessary to take into account penetration of the compensating field behind the grid, but this will only introduce an insignificant correction. Substituting (5) into (4), we obtain:

$$V^+ = N \cdot e \cdot \Delta U \cdot R \cdot \cos\Theta / d \cdot C \cdot U. \qquad (6)$$

Thus, to determine the $V^+$ potential arising on the anode under the action of positive ions, it is only necessary to determine the value of the compensating voltage $\Delta U$. The accuracy of the described method is mainly determined by the accuracy of establishing the moment of compensation. To increase the sensitivity, it is better to use the extra-pure argon as the filling gas, since this gas has the lowest ion-electron recombination probability. Therefore, in this case pulses of considerable amplitude are obtained even in weak fields. In addition, the sensitivity can be increased by increasing the voltage at the anode. More reliable results are obtained when, for determining the moment of the fields compensation one detects pulses from the cathode, since in this case it is possible to trace the change of the sign of the pulses and to determine precisely the moment of their disappearance (Fig. 2). In this way, it is possible to measure the value of $\Delta U / U$ with an accuracy of $(3 - 5)$ %.

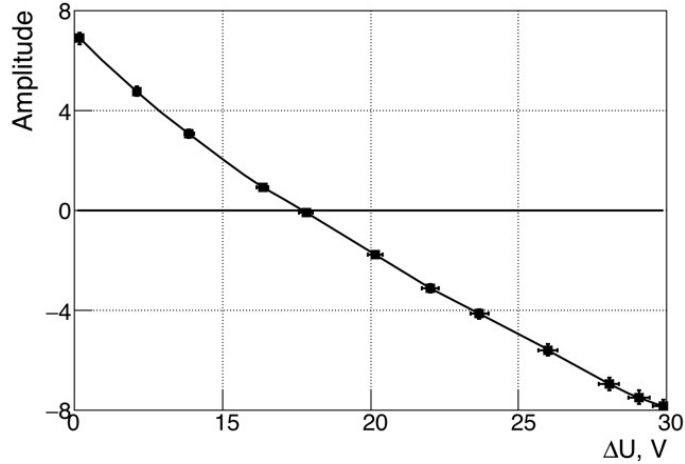

**Fig.2.** *Dependence of the amplitude of the pulses on the cathode (in relative units ) on the value of the compensating voltage. The chamber geometry parameters: anode – grid distance p =58 mm; cathode – grid distance d = 100 mm; grid wire radius r = 0.05 mm; wire spacing S = 1.0 mm. Anode potential U = 2.9 kV. Grid potential zero; Cathode potential $0 + \Delta U$.*

**Comparison of experimental data with the results of the theoretical calculations**

The described above method was used to verify the results of theoretical calculations presented in ref. [1]. According to [1], the grid transparency coefficient $\sigma$ is determined by the following parameters :

$$\sigma = l/(p + l), \qquad (7)$$

$$l = (\rho^2/4 - \ln \rho) \cdot S/2\pi; \quad \rho = 2\pi r/S. \qquad (8)$$

Here $p$ is the distance between the anode and the grid, $r$ is the radius of the grid wires, $S$ is the wire spacing. Using the grid transparency coefficient $\sigma$, one can derive a formula for determination of the potential $V^+$ induced by the positive ions:

$$V^+ = N \cdot e \cdot R \cdot \cos\Theta \cdot \sigma / d \cdot C. \qquad (9)$$

Comparing (9) with (6), we notice that they completely coincide if we put:

$$\Delta U/U = \sigma. \qquad (10)$$

This relation is used to compare the experimental and calculated data. The results of this comparison for two grids are presented in Table 1. The values of $\sigma_{calculated}$ and $\sigma_{exp}$ were determined according to expressions (7-8) and (10), respectively.

**Table 1.** Comparison of the calculated $\sigma_{calculated}$ and measured $\sigma_{exp}$ grid transparency coefficients

| Grid | $r$, mm | $S$, mm | $p$, mm | $\sigma_{calculated}$ | $\sigma_{exp}$ |
|---|---|---|---|---|---|
| 1 | 0,05 | 1,5 | 58 | 0,0065 | 0,0061 |
| 2 | 0,05 | 3,0 | 58 | 0,0185 | 0,0143 |

One can see that the experimental values for the the grid transparency coefficient are slightly lower than the calculated ones, the discrepancy being larger (20%) for the more transparent grid 2. On the other hand, while using the low-transparency grids (such as grid 1), the application of the theoretical expressions (7), (8) is quite acceptable.

In conclusion, the authors express their gratitude to A. P. Komar for his interest to this work and to G. E. Solyakin, who took part in the discussions of the results.

## Post Scriptum

Nowadays, the grid transparency can be calculated using a Garfield package. As an example, we present here the results of such calculations for an ionization chamber with the following geometry parameters: anode – grid distance $p = 10$ mm; cathode – grid distance $d = 30$ mm. The calculations were performed for two values of the wire radius: $r = 25$ μm and $r = 50$ μm. For each radius the wire spacing was varied: $S = 1.0$ mm, $S = 0.5$ mm, $S = 0.25$ mm. The grid transparency coefficient $\sigma$ was determined in the described above Green theorem approach. The electric field $E_d$ in the space between the cathode and the grid was calculated for the anode potential $U = 7500$ V with the cathode and the grid being grounded. The results are shown in Fig.1$_{PS}$.

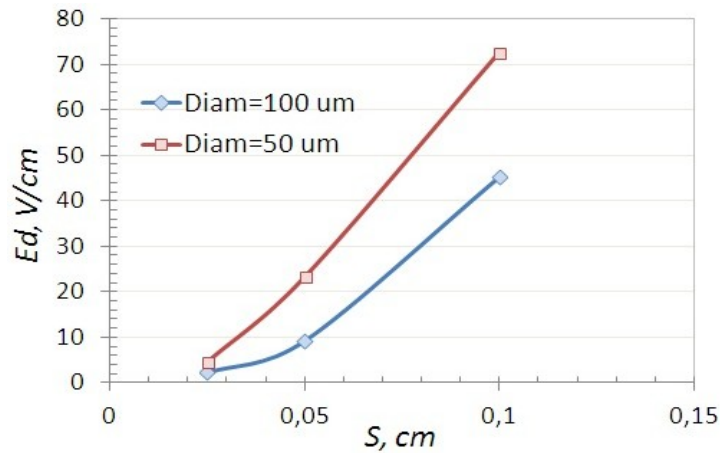

**Fig.1.** The electric field $E_d$ in the space between the cathode and the grid for the anode potential $U = 7500$ V with the cathode and the grid being grounded. The chamber geometry parameters: $p = 10$ mm, $d = 30$ mm. The calculations were performed for two values of the wire radius and three values of the wire spacing. (Acknowledgment to G.Gavrilov who performed these calculations)

With the calculated field $E_d$, the grid transparency was obtained according to expression (10) of this note:

$$\sigma = \Delta U/U = E_d\, d / U \qquad (1_{PS})$$

The results are presented in Table 1 and in Table 2. These results are compared with the results of calculations of the grid transparency with the method of O.Buneman et al. [1] using the expressions (7) and (8).

**Table 1$_{PS}$** Grid transparency calculated using the Garfield package ( $\sigma_{Garfield}$ ) and with the method of O.Buneman et al. ( $\sigma_{Buneman}$ ) for the grid wire radius 25 μm in function of the wire spacing. The chamber geometry parameters : $p$ =10 mm, $d$ = 30 mm.

| Wire spacing $S$ mm | Cathode-grid field $E_d$ V/cm | Grid transparency $\sigma_{Garfield}$ % | Grid transparency $\sigma_{Buneman}$ % |
|---|---|---|---|
| 1 | 72.55 | 2.9 | 3.01 |
| 0.5 | 23.3 | 0.93 | 0.92 |
| 0.25 | 4,72 | 0.19 | 0.19 |

**Table 1$_{PS}$** Grid transparency calculated using the Garfield package ( $\sigma_{Garfield}$ ) and with the method of O.Buneman et al. ( $\sigma_{Buneman}$ ) for the grid wire radius 50 μm in function of the wire spacing. The chamber geometry parameters : $p$ =10 mm, $d$ = 30 mm.

| Wire spacing $S$ mm | Cath-grid field $E_d$ V/cm | Grid transparency $\sigma_{Garfield}$ % | Grid transparency $\sigma_{Buneman}$ % |
|---|---|---|---|
| 1 | 45,18 | 1.81 | 1.84 |
| 0.5 | 9.24 | 0.37 | 0.45 |
| 0.25 | 2.28 | 0.09 | 0.13 |

As it follows from the presented data, the values of grid transparency calculated with the Garfield package and with the method of O. Buneman et al. are quite similar for the considered ionization chamber parameters.

**REFERENCES**

1$_{PS}$. *R. Veenhof* , Garfield – simulation of gaseous detectors.  http://garfield.web.cern.ch/garfield/; Garfield, recent development, Nucl. Instrum. Meth. A 419 (1998) 726-730.